# Boundary Element Method for non-adhesive and adhesive contacts of a coated elastic half-space

Qiang Li[1], Roman Pohrt[1], Iakov A. Lyashenko[1,2], Valentin L. Popov [1,3,4,*]

[1] Berlin University of Technology, 10623 Berlin, Germany
[2] Sumy State University, 40007 Sumy, Ukraine
[3] National Research Tomsk State University, 634050 Tomsk, Russia
[4] National Research Tomsk Polytechnic University, 634050 Tomsk, Russia
[*] Corresponding author: v.popov@tu-berlin.de

## Abstract

We present a new formulation of the Boundary Element Method (BEM) for simulating the non-adhesive and adhesive contact between an indenter of arbitrary shape and an elastic half-space coated with an elastic layer of different material. We use the Fast Fourier Transform based formulation of BEM, while the fundamental solution is determined directly in the Fourier space. Numerical tests are validated by comparison with available asymptotic analytical solutions for axisymmetric flat and spherical indenter shapes.

## 1. Introduction

Layered systems of materials having different mechanical properties have attracted a lot of scientific interest over the last decades [1,2]. A well-chosen coating can improve the structural, mechanical, optical or thermal properties at the surface of a bulk material. Layered structure can be created by ion implantation, vacuum deposition, nanostructure burnishing, laser implantation, and other manufacturing technologies. Coatings a widely used for reducing wear, increasing corrosion resistance, controlling friction, influencing adhesion properties, manipulating thermal conductivity and others. Due to the significant influence of surface layers on mechanical properties, a multitude of experimental techniques have been developed for the characterization of coatings, in particular measuring their elastic properties [3].

Mechanics and contact mechanics of coated systems have been studied intensively [4]. The frictionless adhesive contact between a rigid spherical indenter and an elastic multi-layer coated half-space was investigated in [5] using an integral transform formulation. In [6] an approximate analytical theory was proposed for the contact between a coated half-space and a randomly rough surface. Contacts between flat-ended cylindrical indenters and a flat coated half-space were studied in [7]. Theoretical results were compared with experimental data and will be used in the present paper for verification purposes.

Numerical methods have also been intensively developed to study the contact mechanical and tribological behavior of layered materials. The finite element method is most commonly used. It is very flexible and can be applied for various structures without the restriction of linear material behavior. In contrast to finite element method, the FFT-based Boundary Element Method (BEM) is suitable for all problems with linear materials and geometrical behavior. In part because of its much higher numerical efficiency for contact problems, the BEM evolved to be the standard method in research and development. In [8], the boundary element formulation was presented for a contact of an arbitrary shaped indenter with a homogeneous half-space. Later it was extended to include Johnson-Kendall-Roberts (JKR) type adhesion [9]. The method was validated by available analytical solutions including parabolic contacts (JKR solution), toroidal indenters [10] and flat elliptical indenters [11]. Recently it was also applied to contacts between flat-ended indenters of complicated shape and a flat soft body [12]. In [13], the BEM was developed for contacts with power-law graded materials.

In the present paper we propose a further generalization of BEM for the case of a coated half-space. Several numerical tests will be carried out and the results will be compared with the known analytical solutions.

## 2. FFT-based BEM for layered half-space

We consider a half-space with a single elastic coating of thickness $h$, elastic modulus $E_1$ and Poisson ratio $\nu_1$. The corresponding elastic constants of the half-space are $E_2$ and $\nu_2$ (Fig. 1). The origin of coordinates is placed at the surface of the layer so that the interface between two media is located at $z = h$.

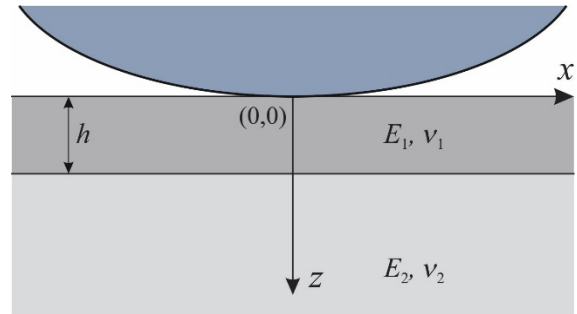

**Fig. 1** Scheme of the system under consideration. An elastic layer with thickness $h$, elastic modulus $E_1$ and Poisson ratio $\nu_1$ is located on top of an elastic half-space with elastic parameters $E_2$ and $\nu_2$.

In previous versions of the BEM for contact of homogeneous and power-law graded materials [8, 13], we proceeded from the fundamental solution in coordinate space and the corresponding integral formulation of the stress-displacement relation. This integral relation has the form of a convolution of the surface pressure distribution with the fundamental solution $\mathbf{U}_0$. This fundamental solution represents the deformation resulting from a single localized normal force. For the numerical solution of the contact problem, we consider a square region on the surface of the body with the size $L \times L$, discretized with $N$ cells in each direction. The size of each of the $N^2$ square cells is $\Delta x = \Delta y = \Delta$. Pressure is assumed to be uniform in each cell (see Fig. 2).

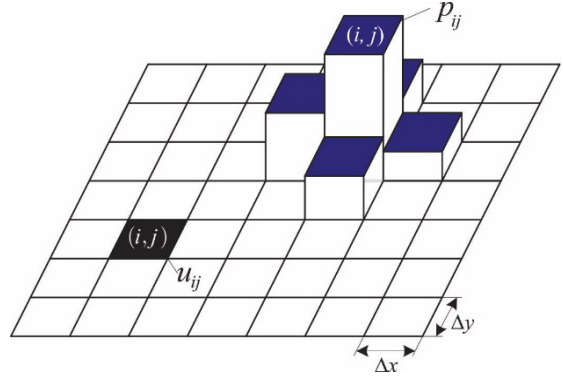

**Fig. 2** Discretization of the simulation area.

In the discretized form, the pressure-displacement relation can be written as

$$\mathbf{u} = \mathbf{K}\mathbf{p}. \tag{1}$$

where $\mathbf{p}$ is stress distribution acting on the surface (vector of the length $N^2$ with values of pressure in the corresponding discrete cells), $\mathbf{u}$ is the normal displacement of surface elements due to applied pressure, and $\mathbf{K}$ is the stiffness matrix having the size $N^4$. The contact problem is solved in BEM iteratively. In each step, the displacements for a given (assumed) pressure distribution are determined by evaluation Eq. (1). Because of the convolution-type integral equation and the resulting structure of matrix $\mathbf{K}$, the operation is performed using direct and inverse Fast Fourier Transforms (FFT):

$$\mathbf{u} = \mathbf{IFFT}\left[\mathbf{FFT}(\mathbf{U}_0)\cdot\mathbf{FFT}(\mathbf{p})\right]. \tag{2}$$

The number of operation for performing this operation is on the order of $o(N^2 \log N)$ (as compared with $o(N^4)$ for direct evaluation of (1).

The stiffness matrix $\mathbf{K}$ is basically a long version of the discretized fundamental solution of the problem. To be used in (2), a known fundamental solution $\mathbf{U}_0$ must first be Fourier-transformed. However, if the fundamental solution can be found directly in the Fourier-space, this step could be skipped. In the present paper, we determine the fundamental solution for a layered system directly in the Fourier-space in analytic form thus saving both one Fourier transform and memory space.

For the derivation of the fundamental solution in Fourier-space, we consider a harmonic pressure distribution acting on the surface of the layer with wave vector $\mathbf{k}$ and amplitude $p_0$:

$$p = p_0 e^{i\mathbf{k}\mathbf{r}}. \tag{3}$$

Here $\mathbf{r}$ is the (two-dimensional) radius-vector in the contact plane.

For the homogeneous case with $E_1 = E_2 = E$ and $\nu_1 = \nu_2 = \nu$, the displacement at the surface ($z = 0$) is equal to [14]

$$u_z(\mathbf{k}) = \frac{2 p_0}{E^* |\mathbf{k}|}, \tag{4}$$

where $E^*$ is the reduced elastic modulus $E^* = E/(1-v^2)$. In the following, we provide the corresponding solution for a layered system.

The equilibrium equation of an elastic isotropic medium reads [14]:

$$\text{grad div }\mathbf{u} + (1-2v_{1,2})\Delta\mathbf{u} = 0, \tag{5}$$

where the Poisson-number has to be chosen $v_1$ for the layer and $v_2$ for the half-space. The displacement vector $\mathbf{u}$ will also have the form of a plane wave with the same wave vector $\mathbf{k}$:

$$\mathbf{u} = u_k^0(z)e^{ikr}\mathbf{e}_k + u_z^0(z)e^{ikr}\mathbf{e}_z. \tag{6}$$

Here $\mathbf{e}_k$ and $\mathbf{e}_z$ are unit vectors in directions of the wave vector and perpendicular to the contact plane correspondingly. Symbols $u_k$ and $u_z$ are projections of the displacement vector on the corresponding directions. The amplitudes $u_k^0$ and $u_z^0$ are functions of the vertical coordinate $z$ only. By substitution of (6) in (5) we obtain a system of two differential equations of second order

$$\frac{\partial^2 u_k^0(z)}{\partial z^2} + \frac{ik}{1-2v_{1,2}}\frac{\partial u_z^0(z)}{\partial z} - \frac{2(1-v_{1,2})k^2}{1-2v_{1,2}}u_k^0(z) = 0 \tag{7}$$

$$\frac{\partial^2 u_z^0(z)}{\partial z^2} - \frac{ik}{2(v_{1,2}-1)}\frac{\partial u_k^0(z)}{\partial z} + \frac{(1-2v_{1,2})k^2}{2(v_{1,2}-1)}u_z^0(z) = 0. \tag{8}$$

These equations should be completed through the following boundary conditions:

(a) Continuity of displacements at the interface between the half-space and coating.

(b) Vanishing of tangential stresses at the contact plane (frictionless problem).

(c) Given normal stress distribution at the surface, Eq. (3): $\sigma_{zz}(r, z=0) = -p_0 e^{ikr}$.

(d) Vanishing of stresses and displacements at infinite depth.

For the plain normal contact problem, we only need normal displacements at the contact surface. Simple but cumbersome calculations lead to the following result:

$$u_z(r, z=0) = \frac{2p_0(1-v_1^2)(Ae^{-4kh} + Bkhe^{-2kh} + D)}{kE_1(-Ae^{-4kh} - Bk^2h^2e^{-2kh} + 2Ce^{-2kh} + D)}e^{ikr} \tag{9}$$

with the constants $A, B, C, D$ given by the following expressions:

$$A = \left[E_2(3-4v_1)(1+v_1) - E_1(3-4v_2)(1+v_2)\right] \times \left[E_1(1+v_2) - E_2(1+v_1)\right],$$
$$B = 4\left[E_2(1+v_1) + E_1(3-4v_2)(1+v_2)\right] \times \left[E_1(1+v_2) - E_2(1+v_1)\right],$$
$$C = E_1^2(4v_2-3)(v_2+1)^2 - 2E_1E_2(v_1+1)(2v_1-1)(v_2+1)(2v_2-1) + \tag{10}$$
$$+E_2^2(8v_1^2-12v_1+5)(v_1+1)^2,$$
$$D = \left[E_2(1+v_1) + E_1(3-4v_2)(1+v_2)\right] \times \left[E_2(3-4v_1)(1+v_1) + E_1(1+v_2)\right].$$

For any given, periodic pressure distribution **p**, the vertical displacement of the layer surface can now be calculated explicitly using Eq. (9)

$$\mathbf{u} = \mathbf{IFFT}\left[\frac{2(1-\nu_1^2)}{E_1} \frac{Ae^{-4kh} + Bkhe^{-2kh} + D}{k\left(-Ae^{-4kh} - Bk^2h^2e^{-2kh} + 2Ce^{-2kh} + D\right)} \cdot \mathbf{FFT}(\mathbf{p})\right]. \quad (11)$$

where $k = \sqrt{k_x^2 + k_y^2}$. Similarly to (4), this procedure only gives results for $k \neq 0$, in other words, the pressure distribution must have no DC-component. The usual BEM procedure reduces to performing the FFT of pressure distribution, multiplying the result with the analytical fundamental solution Eq. (9) and performing inverse FFT to find the displacement field. The inverse problem of finding pressure for producing given deformations can be solved by the conjugate graded method as described in [8]. These two steps complete the formulation of BEM for non-adhesive contacts with layered systems. For adhesive contacts, an additional detachment criterion is needed which is discussed in the remainder of this section.

In each step of an adhesive BEM simulation, the pressure distribution is calculated in all discretized grid cells and it has to be decided whether each point remains in contact or detaches. In [9] and [12] it was suggested to make the decision based on the energy balance criterion of Griffith [16]. For a non-periodic system of a homogenous elastic half-space and a rigid indenter, this leads to a local mesh-dependent detachment criterion: A surface element at the boundary of the contact loses its contact as soon as tensile stress in this element exceeds the critical value given by

$$\sigma_c = \sqrt{\frac{E_1^* \Delta \gamma}{0.473201 \cdot \Delta}}. \quad (12)$$

Here $\Delta \gamma$ is the specific work of adhesion between the indenter and substrate, and $E_1^* = E_1 / (1-\nu_1^2)$. Note that this criterion contains only elastic properties of the coating. This criterion applies also to layered systems, as long as the size $\Delta$ of the discrete cell is much smaller than the thickness of the layer. Under this assumption, the elastic energy released due to the detachment of an element is completely "confined" in the coating, thus the detachment criterion has the same form as in the case of the homogeneous material [9,12] with elastic properties of the coating.

The calculation procedure for numerical simulation of an adhesive contact is basically the same as for non-adhesive contacts. The only difference is in the condition for the loss of contact. If a detachment process is considered, then in each step of the detachment, the indenter is moved upwards by a distance $\Delta d$ (displacement-controlled pull-off). First it is assumed that the contact area does not change, so that all displacements of contact points are augmented by a constant increment $\Delta d$. In the second step, the new stress distribution $p'$ is calculated, which satisfies the new displacement field (inverse problem). In the third step, stresses are checked in all elements at the boundary of the contact area. If the tensile stress in an element is larger than the critical value (12), this element detaches (the stress is set zero), and a reduced contact area $A'$ is obtained. The stress distribution is calculated again with the new contact area and the contact criterion is checked again. This iteration procedure is continued until the tensile stress in each element is smaller than $\sigma_c$ (12). Next, the simulation continues with the next pull-off step.

## 3. Numerical results and comparison with analytic solutions

In [17, 18] it was shown that the solution for axis-symmetric adhesive contact problems can be deduced from the solution of the non-adhesive contact problem: The critical separation distance $d_c$ in adhesive contact is determined by the equation

$$\frac{\mathrm{d}k(a)}{\mathrm{d}a}\frac{d_c^2}{2} = 2\pi a \Delta \gamma, \tag{13}$$

where $k(a)$ is the dependency of the incremental stiffness on the contact radius $a$ for the *non-adhesive* contact problem. Let us apply this relation to the limiting case of soft layer bonded to a rigid substrate. The asymptotically exact result in this case with the additional condition $a \gg h$ is given by [17,18, 19]

$$k = \frac{\pi a^2}{h}\tilde{E}_1, \tag{14}$$

$$\tilde{E}_1 = E_1 \frac{1-\nu_1}{(1+\nu_1)(1-2\nu_1)}. \tag{15}$$

Substitution into (13) provides the following critical values for the indentation depth and the pull-off force.

$$d_c = -\sqrt{\frac{2\Delta\gamma h}{\tilde{E}_1}}, \quad F_c = kd_c = -\pi a^2 \sqrt{\frac{2\tilde{E}_1 \Delta \gamma}{h}}. \tag{16}$$

In the frame of the proposed BEM formulation, we can simulate this limiting case by assuming a very large ratio of $E_2/E_1$. Simulation results of the pull-off of a flat cylindrical indenter are shown in Fig. 3 for two ratios $E_2/E_1 = 10^5$ and $E_2/E_1 = 10^2$ with $E_1 = 2 \cdot 10^9$ Pa, $\nu_1 = \nu_2 = 0.3$ and $a = 50h$. The displacement and the force are normalized to the critical values (16): $\tilde{F} = F/|F_c|$ vs $\tilde{d} = d/|d_c|$. As expected, the force-displacement relation is linear up to the moment of sudden complete detachment. In the case of $E_2/E_1 = 10^2$, this instability point does not match the critical values (16) $\tilde{F} = -1$ and $\tilde{d} = -1$. This is due to the fact that the two ratios $E_2/E_1 \approx a/h$ are of the same order of magnitude, violating the requirements of the asymptotic solution. In contrast for $E_2/E_1 = 10^5$ the critical values approach the point $(-1,-1)$ very closely thus validating the correctness of the used criterion.

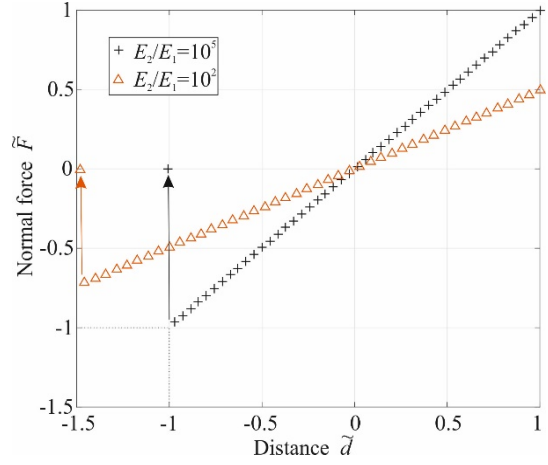

**Fig. 3** Dependencies of dimensionless elastic force $\tilde{F}$ vs. dimensionless distance $\tilde{d}$ for adhesive contact of a cylindrical indenter on an elastic layer.

## 3.1 Indentation by a parabolic indenter

For a thin elastic layer, an asymptotically exact analytic solution exists for arbitrary indenter shapes provided the condition $a \gg h$ remains fulfilled. In the limiting case $E_2 \to \infty$, displayed in Fig. 4, the solution for a parabolic indenter reads [18,19]:

$$F = \frac{\pi \tilde{E}_1 a^2}{h}\left( \frac{a^2}{4R} - \sqrt{\frac{2h\Delta\gamma}{\tilde{E}_1}} \right), \tag{17}$$

$$d = \frac{a^2}{2R} - \sqrt{\frac{2h\Delta\gamma}{\tilde{E}_1}}. \tag{18}$$

The critical values of force, separation and contact radius are given by

$$d_{crit} = -\sqrt{\frac{2h\Delta\gamma}{\tilde{E}_1}}, \; a_{crit} = \left(\frac{8R^2 h\Delta\gamma}{\tilde{E}_1}\right)^{1/4}, \; F_{crit} = -2\pi R\Delta\gamma. \tag{19}$$

With dimensionless variables $\tilde{a} = \dfrac{a}{a_{crit}}$, $\tilde{d} = \dfrac{d}{|d_{crit}|}$, $\tilde{F} = \dfrac{F}{|F_{crit}|}$, the dependencies of the normal force on indentation depth and contact radius (17) and (18) can be written in the form

$$\tilde{F} = \tilde{a}^4 - 2\tilde{a}^2, \tag{20}$$

$$\tilde{d} = \tilde{a}^2 - 1. \tag{21}$$

These relations are plotted with a black dashed line in Fig. 5 (a) and (b). The corresponding numerical results are shown by the curve with cross symbols for the case of $\alpha = 15$. The other parameters can be found in the following discussion. One can see that it very close to the analytical ones. The small discrepancy may be due to finite values of $E_2/E_1$ and $a/h$ used in the numerical simulation.

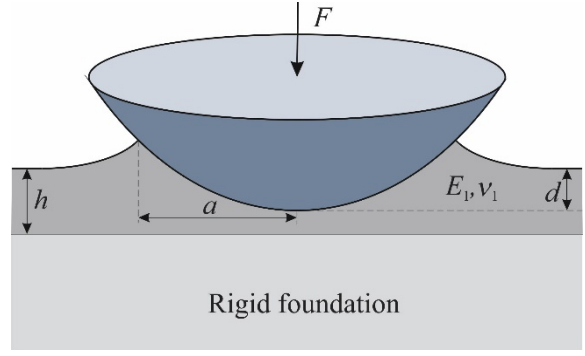

**Fig. 4** Adhesive contact between a rigid indenter and an elastic layer bounded to a rigid foundation.

In the opposite limiting case of the contact radius being small compared to the thickness of the layer, analytical solutions exist in form of asymptotic series in dimensionless parameter $\varepsilon = a/h \ll 1$ [20]:

$$F = \frac{4E_1^* a^3}{3R}\left(1-\varepsilon^3 \frac{8a_1}{3\pi}\right)\left(1-\frac{3R\sqrt{2\pi E_1^* a \Delta\gamma}}{2E_1^* a^2}\right), \quad (22)$$

$$d = \frac{a^2}{R}\left[1-\varepsilon\frac{4a_0}{3\pi}-\varepsilon^3\frac{16a_1}{5\pi}+\varepsilon^4\frac{32a_0 a_1}{9\pi^2} - \right.$$
$$\left. -\frac{R\sqrt{2\pi E_1^* a \Delta\gamma}}{E_1^* a^2}\left(1-\varepsilon\frac{2a_0}{\pi}-\varepsilon^3\frac{16a_1}{3\pi}+\varepsilon^4\frac{16a_0 a_1}{3\pi^2}\right)\right], \quad (23)$$

In the case of a rigid foundation $E_2 \to \infty$, the coefficients in (22) and (23) are given by

$$a_m = \frac{(-1)^m}{2^{2m}(m!)^2}\int_0^\infty \Lambda(u)u^{2m}\mathrm{d}u, \quad (24)$$

$$\Lambda(u) = \frac{2Le^{-4u}-(L^2+1+4u+4u^2)e^{-2u}}{L-(L^2+1+4u^2)e^{-2u}+Le^{-4u}}, \quad L = 4\nu-3. \quad (25)$$

The dependencies (22) and (23) depend on the following adhesion parameter [21]:

$$\alpha = \sqrt{\frac{2\Delta\gamma R^2}{E_1^* h^3}}. \quad (26)$$

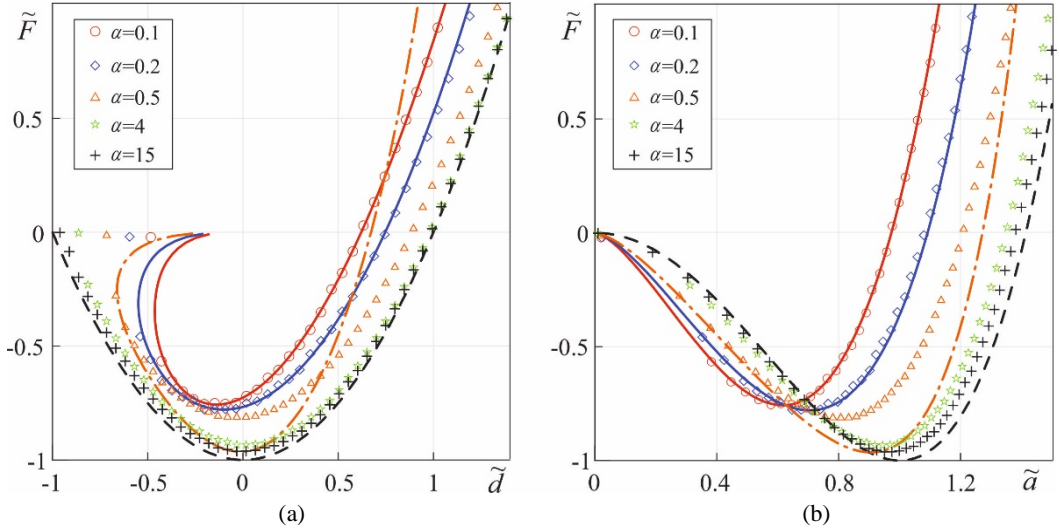

**Fig. 5** Dependencies of dimensionless contact force $\tilde{F}$ on dimensionless indentation depth $\tilde{d}$ and contact radius $\tilde{a}$, for the adhesive indentation of a spherical indenter into a layered counter body. Curves for different values of adhesion parameter $\alpha$ are shown. Dashed lines depict the dependencies (20) and (21) for the case of $a \gg h$. Dash-dot and solid lines are given by expressions (22) and (23) for the case of $a \ll h$. Symbols are numerical results obtained by the BEM presented in this paper.

Numerically we carried out the pull-off simulation with 5 different adhesion parameters $\alpha$ ranging from 0.1 to 15, where the constant parameters are set as $E_1 = 10^9$ Pa, $E_2 = e^{100}$ Pa (which means $\infty$), $\nu_1 = 0.3$, $h = 2$ mm, $\Delta\gamma = 100$ J/m$^2$ and $\alpha$ is varied by changing the radius of curvature $R$ of the indenter. The results are shown in Fig. 5 in the same dimensionless coordinates as given by Eqs. (20) and (21). The curves for $\alpha = 0.1$ and 0.2 corresponding to small contact radii are compared with the asymptotic relations (22), (23) while that of large parameter $\alpha$ are compared with the asymptotic relations (20) and (21). In both limiting cases we see very good agreement between numerical and analytical results. For intermediate values of $\alpha$ an interesting behavior can be observed. For example in the case of $\alpha = 0.5$, for the small indentation depth when the contact radius is much smaller than the layer thickness $a \ll h$, a good coincidence can be observed between numerical (triangles) and analytical results (dash-dot line). With an increasing indentation depth, the analytical solution is not valid any more. At large indentation depths, the numerical results approach the dashed line (the other limiting case $a \gg h$) due to a large contact radius. We thus conclude that numerical results coincide with all available analytical results in region of their validity.

## Conclusion

We generalized the boundary element method proposed in [8,9] for contacts with a half-space coated with an elastic layer. With the suggested BEM formulation, we carried out simulations of adhesive contacts with cylindrical flat-ended and parabolic indenters and compared the results with available asymptotic analytic solutions. We found that numerical results coincide with all available analytical results in region of their validity.

# Acknowledgment

Authors acknowledge financial support of the Deutsche Forschungsgemeinshaft (DFG PO 810-55-1) and the German ministry for research and education BMBF, grant No. 13NKE011A. This research was also partially supported by the "Tomsk State University competitiveness improvement program".